\documentclass[twocolumn]{aastex62}

\usepackage{natbib}
\usepackage{amsmath}

\begin{document}
\def\vec#1{\mbox{\boldmath $#1$}}
\def\rev#1{\textcolor{magenta}{{\bf #1}}}
\def\refp#1{\citep{#1}}
\def\reft#1{\citet{#1}}
\newcommand{\average}[1]{\ensuremath{\langle#1\rangle} }

	\title{Frequency-dependent Alfv\'en-wave propagation in the solar wind: \\
	Onset and suppression of parametric decay instability}

	\author[0000-0002-7136-8190]{Munehito Shoda}
	\affiliation{Department of Earth and Planetary Science, The University of Tokyo, Hongo, Bunkyo-ku, Tokyo, 113-0033, Japan }
	\author[0000-0001-5457-4999]{Takaaki Yokoyama}
	\affiliation{Department of Earth and Planetary Science, The University of Tokyo, Hongo, Bunkyo-ku, Tokyo, 113-0033, Japan }
	\author{Takeru K. Suzuki}
	\affiliation{School of Arts \& Sciences, The University of Tokyo, 3-8-1, Komaba, Meguro, Tokyo, 153-8902, Japan}
	
	\correspondingauthor{Munehito Shoda}
	\email{shoda@eps.s.u-tokyo.ac.jp}
 
        \begin{abstract}

	Using numerical simulations we investigate the onset and suppression of parametric decay instability (PDI) in the solar wind, 
	focusing on the suppression effect by the wind acceleration and expansion.
	Wave propagation and dissipation from the coronal base to $1 {\rm \ au}$ is solved numerically in a self-consistent manner;
	we take into account the feedback of wave energy and pressure in the background.
	Monochromatic waves with various injection frequencies $f_0$ are injected to discuss the suppression of PDI, 
	while broadband waves are applied to compare the numerical results with observation.
	We find that high-frequency ($f_0 \gtrsim10^{-3} {\rm \ Hz}$) Alfv\'en waves are subject to PDI.
	Meanwhile, the maximum growth rate of the PDI of low-frequency ($f_0 \lesssim10^{-4} {\rm \ Hz}$) Alfv\'en waves becomes negative due to acceleration and expansion effects.
	Medium-frequency ($f_0 \approx 10^{-3.5} {\rm \ Hz}$) Alfv\'en waves have a positive growth rate but do not show the signature of PDI up to 1 au because the growth rate is too small.
	The medium-frequency waves experience neither PDI nor reflection so they propagate through the solar wind most efficiently. 
        The solar wind is shown to possess frequency-filtering mechanism with respect to Alfv\'en waves.
	The simulations with broadband waves indicate that the observed trend of the density fluctuation is well explained by the evolution of PDI while
	the observed cross-helicity evolution is in agreement with low-frequency wave propagation.
	
	\end{abstract}


\section{Introduction}
	
	It is widely accepted that Alfv\'en waves \citep{Alfve42} play an important role in the heating \citep{Alfve47,Oster61,Matth99} and acceleration \citep{Belch71b,Jacqu77,Heine80} of the solar wind.
	Indeed, Alfv\'en waves are observed in the solar atmosphere \citep{DePon07a,Tomcz07,McInt11,Sriva17} and solar wind \citep{Colem68,Belch71a}.
	Nonthermal line width \citep{Baner09,Hahn013} and Faraday-rotation fluctuations \citep{Hollw82c,Hollw10} also indicate the existence of Alfv\'en waves in the corona.
	Meanwhile, the dissipation process of Alfv\'en waves in the corona and solar wind is still under discussion.
	Since the amount and location of Alfv\'en-wave dissipation vary with respect to the mechanism and strongly affect the coronal temperature and wind velocity \refp{Hanst12}, 
	clarifying the elemental processes is important not only for plasma physics but also for space weather.
	
	There are several processes of Alfv\'en-wave dissipation.
	If there are counter-propagating Alfv\'en waves, Alfv\'en-wave turbulence \refp{Irosh64,Kraic65,Dobro80,Goldr95} evolves.
	In the corona and solar wind, because of the inhomogeneity, 
	Alfv\'en waves partially reflect \refp{Ferra58,Heine80,An00090,Velli93,Cranm05} and Alfv\'en-wave turbulence is sustained \refp{Dmitr03, Ought06}.
	This reflection-driven Alfv\'en-wave turbulence is frequently studied \refp{Matth99,Dmitr02,Verdi07,Perez13,Balle16},
	and some models explain the heating and acceleration of the solar wind self-consistently \citep{Cranm07,Verdi10}.
	Alfv\'en-wave turbulence is also important for the energy cascade and the formation of the power spectrum \refp{Verdi12a,Balle17}.
	
	When the Alfv\'en velocity is inhomogeneous perpendicular to the magnetic field lines, phase mixing begins \refp{Heyva83,DeGro02,Goose12}.
	The density variation across the magnetic field lines is observed in the corona \refp{Tian011,Raymo14}, and this indicates the possibility of phase mixing.
	Several studies show the role of phase mixing and related phenomena in the solar atmosphere \refp{Antol15,Kanek15}.
	Recently,  it was numerically shown that phase mixing can generate turbulent structure \refp{Magya17}.
	
	Since the amplitude of an Alfv\'en wave is not small and the plasma beta is low \refp{Gary001,Iwai014,Bourd17}, 
	the (extended) corona and solar wind are preferable locations for the development of parametric decay instability (PDI).
	PDI is a type of instability of an Alfv\'en wave \refp{Galee63,Sagde69,Golds78,Derby78} and was recently observed in laboratory plasma \refp{Dorfm16} and in the solar wind \refp{Bowen18}.
	As a result of PDI, a large-amplitude longitudinal wave is generated \refp{Hoshi89,DelZa01}, and the plasma is heated up by the resultant shock wave.
	\reft{Suzuk05,Suzuk06a} demonstrated that, without Alfv\'en-wave turbulence, the coronal heating and solar-wind acceleration are explained self-consistently by PDI.
	These studies were extended to two dimensions (2D) by \reft{Matsu12,Matsu14}.
	In addition, the cross-helicity evolution in the fast solar wind \refp{Bavas82,Bavas00} might be due to PDI \citep{Malar96,Malar00,Shoda16}.
	\reft{Chand18} also argued that the $1/f$ spectrum observed in the fast solar wind \refp{Bruno13} possibly results from PDI.
	
	We note that Alfv\'en-wave turbulence and PDI are not independent of each other,
	because PDI generates large-amplitude backscattered Alfv\'en waves \refp{Sagde69,Golds78} and enhances the heating by Alfv\'en-wave turbulence.
	\citet{Shoda18a} showed that, 
	due to PDI, the turbulence heating rate per unit mass increases ($\sim 10^{11} {\rm \ erg \ g^{-1} \ s^{-1}}$) 
	compared with the reduced- magnetohydrodynamic MHD (without-PDI) value ($\sim 10^{10} {\rm \ erg \ g^{-1} \ s^{-1}}$) \citep{Perez13,Balle16}.
	
	Amongst the aforementioned dissipation processes, we focus on PDI in this study.
	The PDI of monochromatic Alfv\'en waves in a time-independent, uniform background with MHD approximation is well studied.
	In the limit of $\beta \ll 1$ and $\eta = \delta B / B_0 \ll 1$ where $B_0$ and $\delta B$ denote the mean and fluctuating magnetic field, respectively, the growth rate is given as \citep{Galee63,Sagde69}
	\begin{align}
		\gamma / \omega_0 = \frac{1}{2} \eta \beta^{-1/4}, \label{eq:gamma_sagdeev}
	\end{align}
	where $\omega_0$ is the angular frequency of the parent wave.
	Here we define $\beta$ as
	\begin{align}
		\beta = c_s^2/v_A^2,
	\end{align}
	where $c_s$ and $v_A$ denote the sound and Alfv\'en speed, respectively.
	The general dispersion relation that considers full four-wave interaction \refp{Lashm76} is given by \reft{Golds78} and \reft{Derby78} as
	\begin{align}
		\left( \omega-k \right) \left( \omega^2 - \beta k^2 \right) \left[ \left(\omega+k \right)^2 -4 \right] \nonumber \\
		= \eta^2 k^2 \left( \omega^3 + k \omega^2 - 3\omega + k \right),
		\label{eq:gamma_gd78}
	\end{align}
	where $\omega$ and $k$ are normalized by the parent-wave frequency $\omega_0$ and wavenumber $k_0$.
	In this study, we call Eq. (\ref{eq:gamma_gd78}) the Goldstein--Derby dispersion relation.
	By solving Eq. (\ref{eq:gamma_gd78}), \reft{Golds78} confirmed that 
	the classical understanding that the parent wave decays into a forward acoustic wave and a backward Alfv\'en wave is correct in the low-beta regime.
	In the high-beta plasma, however, the behavior of the instability changes \refp{Jayan93}.
	The linear stage of this {\it ideal} (monochromatic, time-independent, and uniform) case is well understood.
	The nonlinear stage of PDI is also frequently studied using numerical simulation.
	\reft{Hoshi89} studied the linear-to-nonlinear evolution of PDI.
	This study was extended to multi-dimensional simulations in both low- and high-beta cases \refp{Ghosh94a,Ghosh94b}.
	\reft{DelZa01} investigated the evolution of PDI with different plasma parameters, different dimensions and different boundary conditions to show the robustness of PDI.
	Recently, the three-dimensional (3D) hybrid simulation of PDI-driven turbulence has been studied \refp{Fu00017}.

	There are several studies on the linear growth rate of PDI under {\it non-ideal} situations.
	Two-fluid and kinetic simulations were performed \refp{Teras86,Nariy08} 
	The PDI of non-monochromatic Alfv\'en waves tends to have a smaller growth rate \refp{Cohen74b,Umeki92,Malar96,Malar00}.
	If the background is turbulent, the growth rate is quenched compared with the {\it ideal} value \refp{Shi0017}.
	The solar wind acceleration and expansion also work to reduce the growth rate \refp{Tener13,DelZa15}.
	Recently the effect of temperature anisotropy on PDI has also been also studied \refp{Tener17b}.
	
	Specifically in the solar wind close to the Sun, wind acceleration and expansion play an important role.
	Such effects are frequently studied using a local co-moving box in the so-called accelerating expanding box (AEB) model \refp{Velli92,Grapp93,Grapp96,Tener17a}.
	One problem with the AEB model is that the dynamics and energetics are not self-consistent; 
	initially, we have to assume the background quantities such as flow speed or Alfv\'en speed and ignore the feedback of wave heating and acceleration on them.
	Our motivation is to test the idea obtained from the AEB model using a non-local simulation box that extends from the corona to the distant heliosphere.
		
	This paper is organized as follows.
	In Section \ref{sec:method}, we describe the basic equations, numerical scheme, and boundary conditions used in this study.
	Section \ref{sec:mono} and Section \ref{sec:broad} describe the results with monochromatic wave injection and broadband wave injection, respectively. 
	We summarize this paper in Section \ref{sec:summary}

\section{Numerical method } \label{sec:method}

\subsection{Basic equations and setting}

	We used the same equations as those in \reft{Shoda18a} and
	considered a one-dimensional system whose coordinate $r$ is curved along the background magnetic field line.
	The basic equations used were
	\begin{align}
		&\frac{\partial}{\partial t} \left( \rho r^2 f \right) + \frac{\partial }{\partial r} \left( \rho v_r r^2 f \right) =0,  \label{eq:mass} \\ 
    		&\frac{\partial}{\partial t} \left( \rho v_r r^2 f \right) + \frac{\partial }{\partial r} \left[ \left( \rho {v_r}^2 + p + \frac{{\vec{B}_{\perp}}^2}{8\pi} \right) r^2 f \right] \nonumber \\
    		&= \left( p + \frac{\rho {\vec{v}_{\perp}}^2}{2} \right) \frac{d}{dr} \left( r^2 f \right) - \rho g r^2 f, \label{eq:eomradial} \\
    		&\frac{\partial}{\partial t} \left( \rho \vec{v}_{\perp} r^3 f^{3/2} \right) + \frac{\partial }{\partial r} \left[ \left( \rho v_r \vec{v}_{\perp} - \frac{B_r \vec{B}_{\perp}}{4 \pi} \right) r^3 f^{3/2} \right] \nonumber \\
   		&= -\hat{\vec{\eta}}_1 \cdot \rho \vec{v}_{\perp} r^3 f^{3/2} - \hat{\vec{\eta}}_2 \cdot \sqrt{\frac{\rho}{4 \pi}} \vec{B}_{\perp} r^3 f^{3/2}, \label{eq:vt} \\
    		&\frac{\partial}{\partial t} \left( \vec{B}_{\perp} r \sqrt{f} \right) + \frac{\partial }{\partial r} \left[ \left( \vec{B}_{\perp} v_r - B_r \vec{v}_{\perp} \right) r \sqrt{f} \right] \nonumber \\
		&= -\hat{\vec{\eta}}_1 \cdot \vec{B}_{\perp} r \sqrt{f} - \hat{\vec{\eta}}_2 \cdot \sqrt{4 \pi \rho} \vec{v}_{\perp} r \sqrt{f}, \label{eq:bt} \\
   	 	&\frac{d}{dr} \left(B_r  r^2 f  \right) = 0, \label{eq:divb} \\
  		&\frac{\partial}{\partial t} \left[ \left( e + \frac{1}{2} \rho \vec{v}^2 + \frac{\vec{B}^2}{8 \pi} \right) r^2 f \right] \nonumber \\
  		&+ \frac{\partial }{\partial r} \left[\left( e + p + \frac{1}{2} \rho \vec{v}^2  + \frac{{\vec{B}_{\perp}}^2}{4 \pi} \right ) v_r r^2 f - B_r \frac{\vec{B}_{\perp} \cdot \vec{v}_{\perp}}{4 \pi} r^2 f \right] \nonumber \\
    		&= r^2 f \left( - \rho g v_r + Q_{\rm cond}  \right), \label{eq:energy} \\
		& e = \frac{p}{\Gamma-1}, \ \ \ \ p = \frac{\rho k_B T}{\mu}. \label{eq:eos} \
	\end{align}
	See Appendix in \reft{Shoda18b} for the derivation.
	We denoted the perpendicular components of $\vec{X}$ as $\vec{X}_\perp = X_x \vec{e}_x + X_y \vec{e}_y$,
	and we assumed that the plasma is composed of only hydrogen and is fully ionized in the entire simulation region.
	Therefore, the mean molecular mass $\mu$ satisfied $\mu = 0.5 m_p$ where $m_p$ is the proton mass.
	$\Gamma$ is the adiabatic specific heat: $\Gamma = 5/3$.
	
	$f$ is the expansion factor of the flux tube \refp{Levin77,Wang090,Arge000}.
	In this study, following \reft{Kopp076} and \reft{Verdi10}, we assumed
	\begin{align}
		f(r) = \frac{f_{\rm exp} \exp \left[ \left( r - r_f \right) / \sigma \right] + f_1 }{\exp \left[ \left( r - r_f \right) / \sigma \right] + f_1},
	\end{align}
	where $f_1 = 1- f_{\rm exp} \exp \left[ \left( R_\odot - r_f \right) / \sigma \right]$, $f_{\rm exp}=10$, $r_f = 1.3 R_\odot$ and $\sigma = 0.5 R_\odot$.
		
	$\hat{\vec{\eta}}_1$ and $\hat{\vec{\eta}}_2$ are coefficient tensors that represent phenomenological turbulent decay.
	\begin{align}
          	&\hat{\vec{\eta}}_1= 
          	\frac{c_d}{4 \lambda} \left(
          		\begin{array}{cc}
            			  |z^+_x| + |z^-_x|  & 0 \\
            			0 &  |z^+_y| + |z^-_y|  \\
          		\end{array}
          	\right), \label{eq:eta1} \\
          	&\hat{\vec{\eta}}_2=
          	\frac{c_d}{4 \lambda} \left(
          		\begin{array}{cc}
           			 |z^+_x| - |z^-_x|  & 0 \\
            			0 &  |z^+_y| - |z^-_y| \\
         		\end{array}
          	\right), \label{eq:eta2}
        \end{align}
        where $z^{\pm}_{x,y}$ are Els\"asser variables \refp{Elsas50}:
        \begin{align}
        		z^{\pm}_{x,y} = v_{x,y} \mp B_{x,y} / \sqrt{4 \pi \rho}.
        \end{align}
        \reft{Shoda18a} showed that these terms are a natural extension of a widely used phenomenological model of Alfv\'en-wave turbulence \refp{Hossa95,Dmitr02,Verdi07,Chand09a}.
        $c_d = 0.1$ was chosen in this study \refp{Balle17}.
        $\lambda$ is the perpendicular correlation length of turbulence.
        We assumed that the correlation length is proportional to the flux-tube radius:
        \begin{align}
        		\lambda  \propto B_r^{-1/2}.
		\label{eq:lambda}
        \end{align}
        Using the phenomenological turbulence term together with Eq. (\ref{eq:lambda}), 
        both local \refp{Cranm07,Verdi10,Lione14,Shoda18a} and global \refp{Holst14} simulations succeeded in modeling the corona and solar wind.
	The correlation length at the coronal base ($\lambda_0$) is
	\begin{align}
		\lambda_0 = 1{\rm \ Mm}.
	\end{align}
	This is based on the assumption that Alfv\'en waves are generated inside the magnetic patches on the photosphere and propagate upward along the flux tube \refp{Balle11}.
	
	$Q_{\rm cond}$ is the heating by thermal conduction given as
	\begin{align}
		Q_{\rm cond} = - \nabla \cdot \vec{q}_{\rm cond} = - \frac{1}{r^2 f} \frac{\partial}{\partial r} \left( r^2 f q_{\rm cond} \right),
	\end{align}
	where $\vec{q}_{\rm cond}$ is the conductive flux and $q_{\rm cond}$ represents its radial component.
	The conductive flux is a combination of Spitzer-H\"arm flux \refp{Spitz53} and free-streaming flux \refp{Hollw74,Hollw76} given as
	\begin{align}
		q_{\rm cond} = \xi q_{\rm SH} + (1-\xi) q_{\rm FS}, \ \ \ \ \xi = \max \left(1, \frac{\rho}{\rho_{\rm SW}} \right)
	\end{align}
	where $\rho_{\rm SW} = 10^{-21} {\rm g \ cm^{-3}}$ and
	\begin{align}
		&q_{\rm SH} = - \kappa_0 T^{5/2} \frac{\partial}{\partial r} T, \\
		&q_{\rm FS} = \frac{3}{4} \alpha p v_r.
	\end{align}
	In GGS-Gaussian units, $\kappa_0 \approx 10^{-6}$.
	We fixed $\alpha=2$ in this study.
	
	Radiative cooling is ignored because of its small contribution to the coronal energy budget \refp{Matsu14,Balle16}.
	The coronal base is cooled down by keeping the bottom temperature fixed.

\subsection{Numerical scheme and boundary conditions}

	We solved the basic equations (\ref{eq:mass})-(\ref{eq:eos}) from the coronal base ($r=1.014R_\odot$) to 1 au ($r= 215 R_\odot$).
	Furthermore, we applied 50 000 uniform grid points to resolve the computational domain. 
	The Harten-Lax-van Leer-discontinuities (HLLD) approximated Riemann solver \refp{Miyos05} with 2nd-order monotone upstream-centered schemes for conservation law (MUSCL) reconstruction \refp{Leer079} was used to calculate the numerical flux, 
	while the 3rd-order strong stability preserving (SSP) Runge-Kutta method \refp{Shu0088} was used for time integration.

	The free boundary condition was imposed on the boundary at 1 au.
	We confirmed that the boundary condition at 1 au does not affect the calculation because the super-sonic and super-Alfv\'enic solar wind is formed in a quasi-steady state.
	This is why we did not need to apply the transmitting boundary condition \refp{Thomp87,DelZa01,Suzuk06a}.
	As for the lower boundary, the conditions were as follows.
	Here we denoted the lower-boundary values with subscript $0$.
	The mass density $\rho$, temperature $T$, and radial magnetic field strength $B_{r}$ were fixed to
	\begin{align}
		\rho_0 = 8.5 \times 10^{-16} {\rm \ g \ cm^{-3}}, \ \ T_0 = 4 \times 10^5 {\rm \ K}, \ \ B_{r,0} = 10 {\rm \ G}.
	\end{align}
	The quantity $B_{r,0}/f_{\rm exp}$ controls the solar-wind velocity \refp{Suzuk04,Suzuk06b,Fujik15,Revil17}.
	According to \reft{Fujik15}, our setting of ($B_{r,0}/f_{\rm exp} = 1 {\rm \ G}$) approximately corresponds to $650 {\rm \ km \ s^{-1}}$ in terms of asymptotic solar wind velocity.
	
	We applied the free boundary conditions for the radial velocity and inward Els\"asser variable:
	\begin{align}
		\left. \frac{\partial}{\partial r} v_r \right|_0 = 0, \ \ \ \ \left. \frac{\partial}{\partial r} \vec{z}^{-} \right|_0 = 0,
	\end{align}
	where $\vec{z}^{-} = \vec{v}_\perp + \vec{B}_\perp / \sqrt{4 \pi \rho}$. 
	As for the upward Els\"asser variable, $\vec{z}^{+} = \vec{v}_\perp - \vec{B}_\perp / \sqrt{4 \pi \rho}$, 
	we applied monochromatic (Section  \ref{sec:mono}) or broadband (Section \ref{sec:broad}) wave injections.
	In both cases, the root-mean-square value of the transverse velocity $v_{\rm rms,0}$ was fixed to $v_{\rm rms,0} = 32 {\rm \ km \ s^{-1}}$.
	In terms of the upward Els\"asser variable, the root-mean-square value was $z^{+}_{\rm rms,0} = 2 v_{\rm rms,0}$,
	because $\left| z^{+}_{x,y} \right| \gg \left| z^{-}_{x,y} \right|$ at the coronal base and
	\begin{align}
		z^{+}_{\rm rms,0} &= \sqrt{ {z^{+}_x}^2 + {z^{+}_y}^2 } \nonumber \\
					&\simeq \sqrt{ \left( z^{+}_x +  z^{-}_x \right)^2 + \left( z^{+}_y + z^{-}_y \right)^2 } \nonumber \\
					&= \sqrt{ 4 \left( v_x^2 + v_y^2 \right) } = 2 v_{\rm rms,0}.
 	\end{align}
	Note that the injected energy flux $F_0$ was kept constant:
	\begin{align}
		F_0 = \frac{1}{4} \rho {z^{+}_{0}}^2 v_{A,0} = 8.4 \times 10^5 {\rm \ erg \ cm^{-3} \ s^{-1}}.
	\end{align}
	This was larger than the required amount of energy injection required to sustain the solar wind in the open field regions \refp{Withb77}.
	
\section{Monochromatic-wave injection } \label{sec:mono}

	We first applied the monochromatic wave injections with different frequencies to discuss the basic properties.
	The boundary condition of the upward Els\"asser variable was
	\begin{align}
		&z^{+}_{x,0} = 2 v_{\rm rms,0} \sin \left( 2 \pi f_0 t \right), \\
		&z^{+}_{y,0} = 2 v_{\rm rms,0} \cos \left( 2 \pi f_0 t \right),
	\end{align}
	where $f_0$ is the injection frequency.

\subsection{Quasi-steady state } \label{sec:qs_mono}
	
	\begin{figure}[!t]
		\begin{center}	 
 		 \includegraphics[width=70mm]{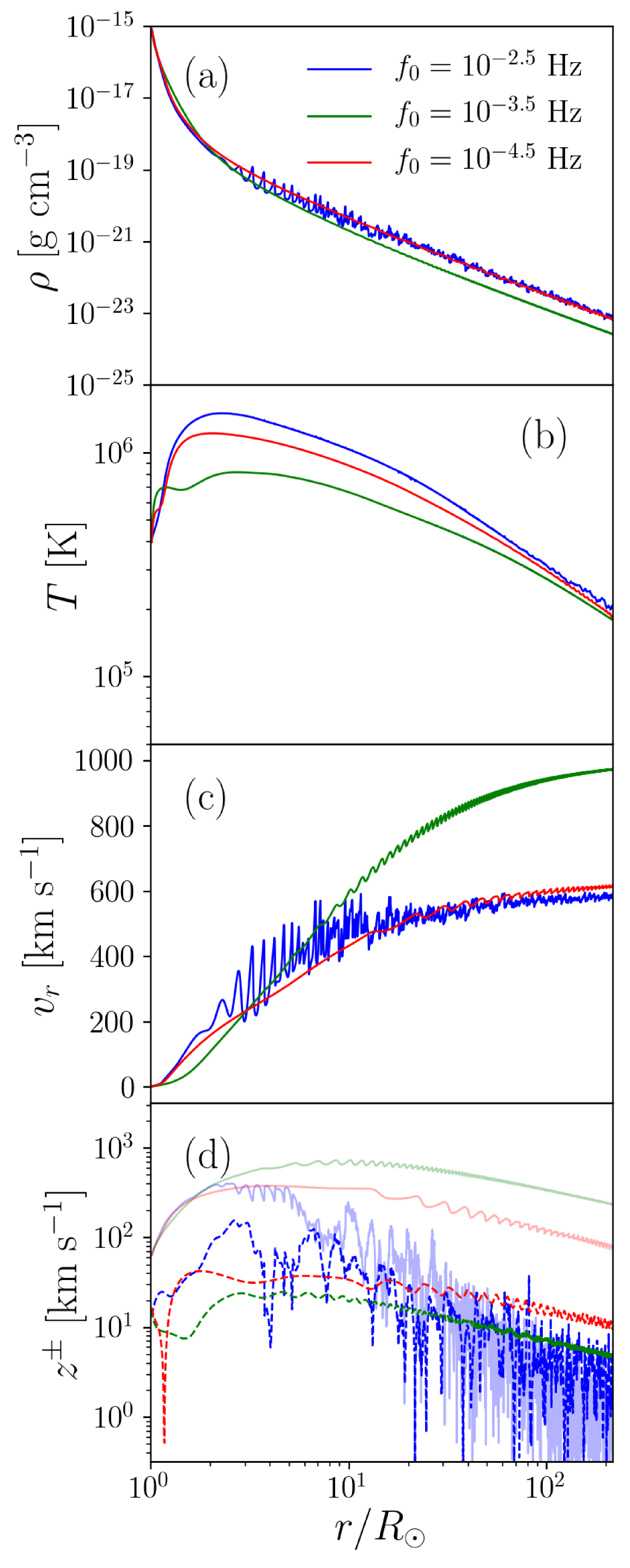} 
  		\end{center}
  		\vspace{-1em}
  		\caption{
				Snapshots of the quasi-steady states with different wave-injection frequencies.
				Blue, green and red lines indicate $f_0 = 10^{-2.5} {\rm \ Hz}$, $10^{-3.5} {\rm \ Hz}$, $10^{-4.5} {\rm \ Hz}$, respectively.
				Panels correspond to a: mass density, b: temperature, c: radial velocity, d: Els\"asser variables.
				In Panel d, transparent and dashed lines indicate $z^{+}$ and $z^{-}$, respectively.
				 }
  		\label{fig:compare_rs_mono}
	\end{figure}
	
	\begin{figure*}[!t]
		\begin{center}	 
 		 \includegraphics[width=180mm]{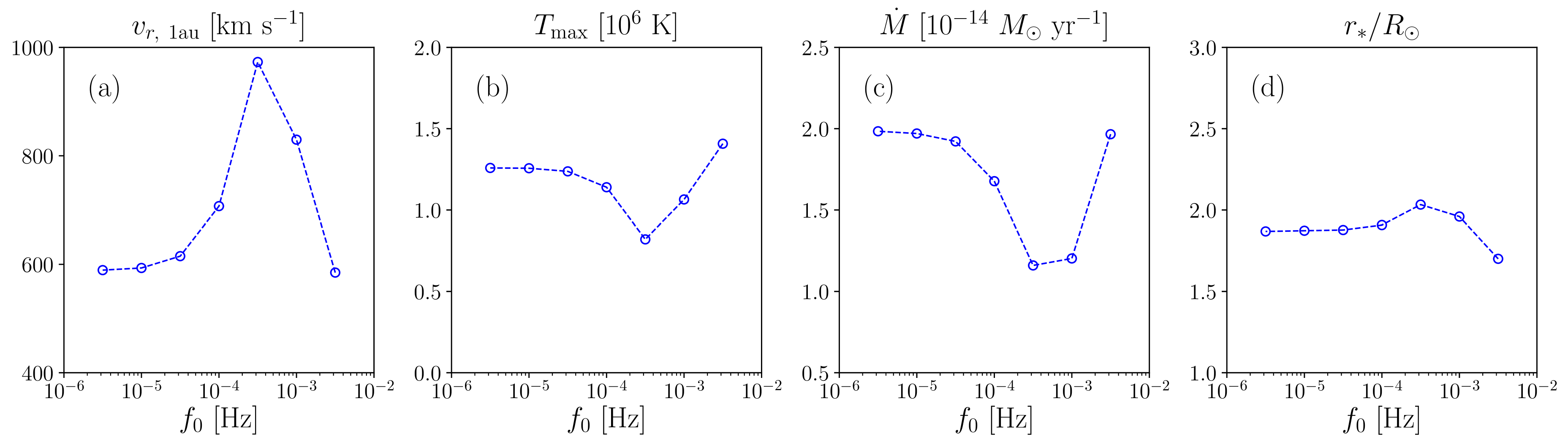} 
  		\end{center}
  		\vspace{-1em}
  		\caption{
				Parameters of the corona and solar wind as functions of wave-driving frequency.
				Panels indicate the a: solar wind speed at 1 au, b: maximum coronal temperature, c: mass-loss rate, and d: sonic point location.
				Each parameter is measured in the time-averaged quasi-steady states. 
  				}
  		\label{fig:fdepend_qs}
	\end{figure*}

	Figure \ref{fig:compare_rs_mono} shows the snapshots of the quasi-steady states of different injection frequencies: 
	$f_0 = 10^{-2.5} {\rm \ Hz \ (blue)}$, $10^{-3.5} {\rm \ Hz \ (green)}$, $10^{-4.5} {\rm \ Hz \ (red)}$. 
	Panels indicate from top to bottom the mass density $\rho$, temperature $T$, radial velocity $v_r$, and Els\"asser variables $z^{\pm} = \sqrt{{z_x^{\pm}}^2 + {z_y^{\pm}}^2}$ (transparent: $z^{+}$, dashed: $z^{-}$).
	
	Although the same amount of energy flux ($F_0 = 8.4 \times 10^5 {\rm \ erg \ cm^{-3} \ s^{-1}}$) was injected in each case, the corresponding quasi-steady states showed different properties.
	Firstly, a significant density fluctuation is observed when $f_0 = 10^{-2.5} {\rm \ Hz}$.
	Because the large density fluctuation is attributed to PDI, it indicates that PDI can develop only when $f_0 > 10^{-3.5} {\rm \ Hz}$.
	Els\"asser variables also show evidence of PDI when $f_0 = 10^{-2.5} {\rm \ Hz}$.
	The ratio $z^{+}/z^{-}$ is smaller than unity partly in $r/R_{\odot}>50$ when $f_0 = 10^{-2.5} {\rm \ Hz}$, while $z^{+}/z^{-} \gg 1$ when $f_0 \lesssim 10^{-3.5} {\rm \ Hz}$.
	A natural interpretation of low $z^{+}/z^{-}$ is that, as a result of PDI, a large amount of reflected Alfv\'en waves is generated \refp{Sagde69,Golds78,Suzuk05} and is advected to 1 au.
	The coronal temperature is the lowest in the medium-frequency case ($f_0=10^{-3.5} {\rm \ Hz}$).
	When $f_0$ is high, because PDI occurs in the sub-Alfv\'enic corona, the coronal plasma is heated up by the shock and turbulence driven by PDI \refp{Shoda18a}.
	However, when $f_0$ is low, Alfv\'en waves reflect efficiently \refp{An00090,Velli93,Cranm05} and the turbulence heating in the corona increases \refp{Matth99,Dmitr02,Ought06}.
	This is why the medium-frequency case, in which PDI does not occur and reflection is weak, shows the lowest temperature of the corona.
	As a result of the lower-temperature corona, the mass density of the wind is smaller and the wind is faster \refp{Hanst12}.
	
	In Figure \ref{fig:fdepend_qs}, we show the dependence of solar wind parameters on $f_0$.
	From left to right, we show the solar-wind velocity $v_r$ at $r=1{\rm \ au}$, maximum temperature $T_{\rm max}$, 
	mass-loss rate $\dot{M} = \rho v_r 4 \pi r^2$, and the sonic point $r_\ast$ where the sound speed $c_s$ is equal to the wind speed $v_r$.
	Here, we assumed $c_s=\sqrt{p/\rho}$ because the plasma is almost isothermal due to the strong thermal conduction near the sonic point.
	Every variable is averaged in time over $10^5 {\rm \ s}$.
	
	Figure \ref{fig:fdepend_qs} shows that the solar wind properties depend non-monotonically on $f_0$; 
	slow, high-temperature, and high-density winds are driven in the cases with high and low $f_0$; 
	in contrast, fast, low-temperature, and low-density winds stream out in the cases with intermediate $f_0$. 
	As explained before, this bimodal behavior can be understood by the different characters of the reflection and dissipation of low- and high-frequency Alfv\'{e}n waves;
	low-frequency waves dissipate by reflection-driven turbulence and high-frequency waves by PDI.
	In addition, Fig. \ref{fig:fdepend_qs} indicates that the corona and solar wind have a frequency-filtering mechanism;
	waves with a medium frequency $f_0 \sim 10^{-3.5} {\rm \ Hz}$ are the least dissipative and most transparent in order to propagate through.
	This might be responsible for the dominance of the hour-scale Alfv\'en waves observed in the solar wind \refp{Belch71a}.   
	
	Some features found in Fig. \ref{fig:fdepend_qs} are consistent with previous research.
	In the high-frequency range, the solar wind velocity (Fig. \ref{fig:fdepend_qs}a) decreases as $f_0$ increases, and this result is consistent with \reft{Ofman98}, 
	who showed the inverse correlation between the injection frequency and the resultant wind speed when $0.35 {\rm \ mHz} \lesssim f_0 \lesssim 3 {\rm \ mHz}$.
	The critical point $r_*$ (Fig. \ref{fig:fdepend_qs}d) has a negative correlation with the temperature.
	This is because the critical point is closer to the Sun when the sound speed is larger \refp{Parke58}.
		
\subsection{Decay law of density fluctuation in the accelerating and expanding solar wind } \label{sec:decaylaw}
	
	Following \reft{Tener13}, we derived the linear decay law for slow magnetoacoustic waves in the accelerating and expanding solar wind.
	We began with the conservation of mass: Eq. (\ref{eq:mass}).
	Assuming that the density and radial velocity have mean $\rho_0$, $v_{r,0}$ and small fluctuation $\delta \rho$, $\delta v_r$ parts, we could express the linearized equation for $\delta \rho$ as
	\begin{align}
		\frac{\partial}{\partial t} \left( \delta \rho S \right) + \frac{\partial}{ \partial r} \left( \rho_0 \delta v_r S \right) +  \frac{\partial}{ \partial r} \left( \delta \rho v_{r,0} S \right) = 0.
		\label{eq:dro_linear}
	\end{align}
	where $S=r^2f$ represents the cross section of flux tube.
	
	We could safely assume that the compressible fluctuations come from upward slow mode because PDI generates the slow-mode wave propagating in the same direction as the parent Alfv\'en wave.
	Therefore, $\delta \rho$ and $\delta v_r$ satisfy a characteristic relation of
	\begin{align}
		\delta \rho / \rho_0 = \delta v_r / c_s.
		\label{eq:characteristic_slow}
	\end{align}
	This relation holds when the slow mode has acoustic nature.
	When $\beta \ll 1$, magnetic and acoustic perturbations decouple with each other.
	In addition, the gravity effect is negligible when the wavelength is much smaller than the scale height of stratification.
	Therefore, Eq. (\ref{eq:characteristic_slow}) is a good approximation because $\beta$ is small and the scale height is large in and above the corona.
		
	Combining Eq. (\ref{eq:dro_linear}) and Eq. (\ref{eq:characteristic_slow}), we had
	\begin{align}
		\frac{\partial}{\partial t} \left( \delta \rho S \right) + \frac{\partial}{ \partial r} \left[ \delta \rho \left( v_{r,0} + c_s \right) S \right] = 0.
	\end{align}
	$\delta \rho$ was assumed to have following form:
	\begin{align}
		\delta \rho \propto \exp \left[ i \left( kr - \omega t \right) \right],
				\label{eq:dro_formulation}
	\end{align}
	From Eqs. (\ref{eq:dro_linear}) and (\ref{eq:dro_formulation}), we have
	\begin{align}
		- i \omega + i k \left( c_s + v_{r,0} \right) + \left( v_{r,0} + c_s \right)  \frac{\partial}{\partial r} \ln S  + \frac{\partial}{\partial r} \left( v_{r,0} + c_s \right) = 0.
	\end{align}
	If the background has little variation and the third term in the left hand side is negligible, the usual dispersion relation of the acoustic wave ($\omega / k = c_s+v_{r,0} $) is obtained.
	If not, we have
	\begin{align}
		\omega = \left( c_s+v_{r,0} \right) k - i \gamma_{\rm acc} - i \gamma_{\rm exp},
	\end{align}
	where 
	\begin{align}
		\gamma_{\rm acc} = \frac{\partial}{\partial r} \left( v_{r,0} + c_s \right),
	\end{align}
	and
	\begin{align}
		\gamma_{\rm exp} = \left( v_{r,0} + c_s \right) \frac{\partial}{\partial r} \ln S,
	\end{align}
	are the damping rates by the acceleration and expansion of the solar wind, respectively. 
	In the linear regime, the density fluctuations have decay rates of $\gamma_{\rm acc} + \gamma_{\rm exp}$.
	
	Since density fluctuation should increase as a result of PDI, acceleration and expansion work to suppress the instability \refp{Tener13,DelZa15}.
	The effective growth rate $\gamma_{\rm eff}$ of PDI is given as
	\begin{align}
		\gamma_{\rm eff} = \gamma_{\rm GD} - \gamma_{\rm acc} - \gamma_{\rm exp},
		\label{eq:effective}
	\end{align}
	where $\gamma_{\rm GD}$ is a growth rate given by the Goldstein--Derby dispersion relation: Eq. (\ref{eq:gamma_gd78}).

	\begin{figure}[!t]
		\begin{center}	 
 		 \includegraphics[width=70mm]{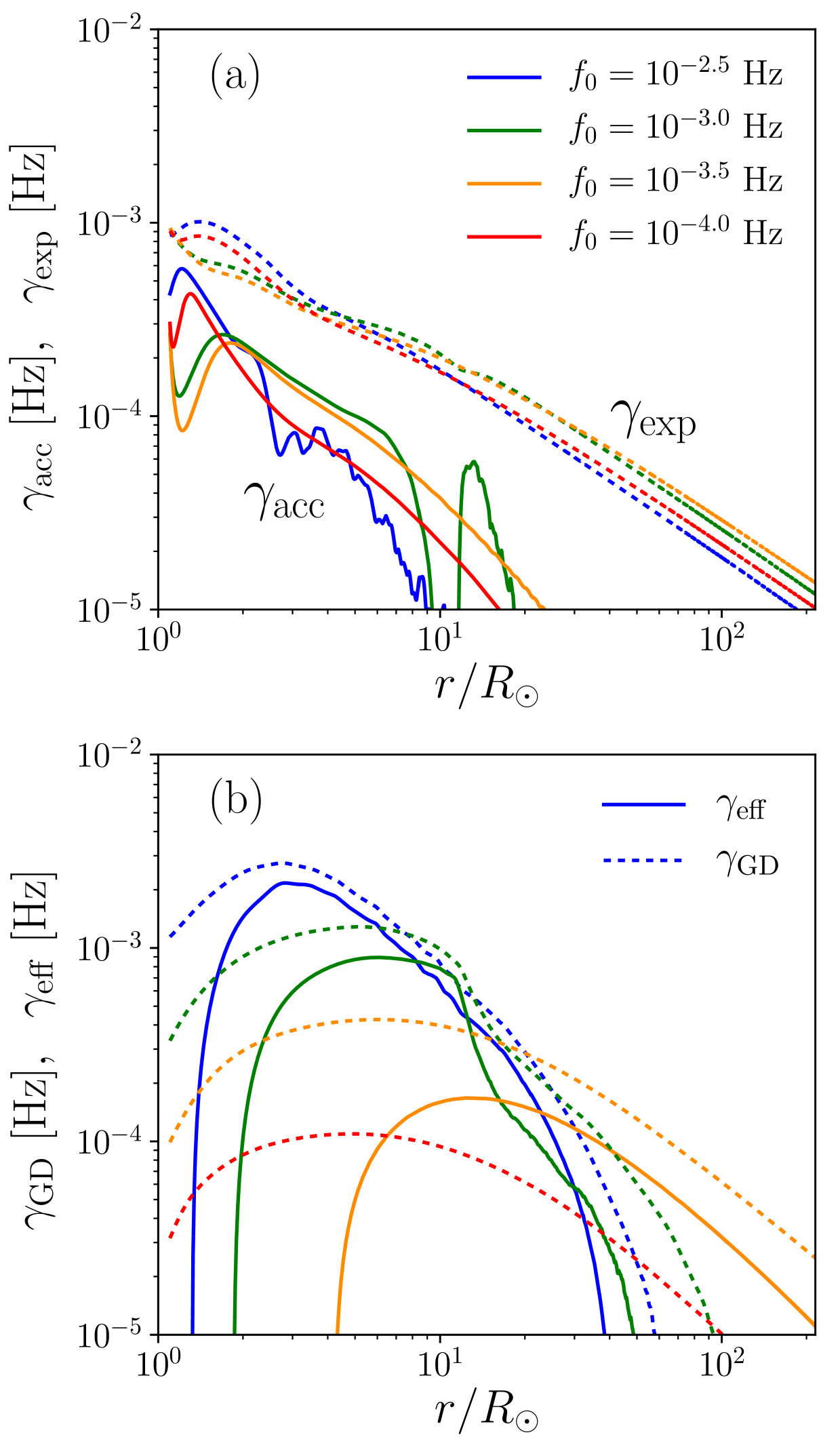} 
  		\end{center}
  		\vspace{-1em}
  		\caption{
				a: Decay rates of density fluctuation due to wind acceleration $\gamma_{\rm acc}$ (solid line) and due to wind expansion $\gamma_{\rm exp}$ (dashed line).
				b: Growth rate of PDI given by the Goldstein--Derby dispersion relation $\gamma_{\rm GD}$ (dashed line) and the effective growth rate $\gamma_{\rm eff}$ (solid line).
				Blue, green, orange, and red lines indicate $f_0 = 10^{-2.5} {\rm \ Hz}$, $10^{-3.0} {\rm \ Hz}$, $10^{-3.5} {\rm \ Hz}$, and $10^{-4.0} {\rm \ Hz}$, respectively.
				We note that $\gamma_{\rm eff}$ of $f_0=10^{-4.0}{\rm \ Hz}$ (red solid) is negative in the entire region.
  				}
  		\label{fig:compare_gamma}
	\end{figure}
	
\subsection{Doppler effect and effective growth rate}

	To discuss the possibility of the onset of PDI for each $f_0$, we calculated $\gamma_{\rm eff}$ using Eq. (\ref{eq:effective}).
	The normalized growth rate $\gamma_{\rm GD} / \omega$ was calculated from Eq. (\ref{eq:gamma_gd78}).
	We should note that $\omega \ne 2 \pi f_0$ because of the Doppler effect by the acceleration of the solar wind.
	$\omega$ should be the intrinsic frequency, that is, the wave frequency observed in a co-moving frame of the solar wind.
	Because the wave frequency observed from a fixed coordinate is constant in a quasi-steady state, the wave number $k(r)$ satisfies
	\begin{align}
		k(r) = \frac{2 \pi f_0}{v_A(r) + v_r (r)}.
	\end{align}
	When we observed this wave in a co-moving frame, the wave number was invariant and the phase speed decreased to $v_A(r)$, and therefore,
	\begin{align}
		\omega = v_A(r) k(r) = 2 \pi f_0 \frac{v_A(r)}{v_A(r) + v_r (r)}.
	\end{align}
	This means that the intrinsic frequency decreases in an accelerating flow.
	In the accelerating expanding box model, this effect is mentioned as the box-stretching effect \refp{Tener17a}.
	A similar argument appears in deriving the wave-action conservation \refp{Dewar70,Heine80}.
	
	We should note that, the wind-acceleration effect appears in different ways.
	As discussed in Section \ref{sec:decaylaw}, the wind acceleration works to reduce the density fluctuation.
	In addition, as discussed here, wind acceleration also causes the Doppler effect.
	
	In Figure \ref{fig:compare_gamma}, we show $\gamma_{\rm eff}$, $\gamma_{\rm GD}$, $\gamma_{\rm acc}$, and $\gamma_{\rm exp}$ as functions of height 
	to see the effects of wind acceleration and expansion on the growth rate of PDI.
	$\gamma_{\rm acc}$ (solid lines) and $\gamma_{\rm exp}$ (dashed lines) are shown in Figure \ref{fig:compare_gamma}a, 
	while $\gamma_{\rm eff}$ and $\gamma_{\rm GD}$ are shown in Figure \ref{fig:compare_gamma}b.
	The colors represent the injection frequency as follows: 
	$f_0=10^{-2.5} {\rm \ Hz}$ (blue), $f_0=10^{-3.0} {\rm \ Hz}$ (green), $f_0=10^{-3.5} {\rm \ Hz}$ (orange), and $f_0=10^{-4.0} {\rm \ Hz}$ (red).
	Figure \ref{fig:compare_gamma}a shows that the expansion ($\gamma_{\rm exp}$) dominates the acceleration ($\gamma_{\rm acc}$) in the damping of the PDI. 
	As a result, $\gamma_{\rm GD}$ is reduced to $\gamma_{\rm eff}$. 
	The reduction factors, $\gamma_{\rm acc}/\gamma_{\rm GD}$ and $\gamma_{\rm exp}/\gamma_{\rm GD}$, are larger for smaller injection frequencies, $f_0$, 
	because $\gamma_{\rm GD}$ is proportional to $f_0$.
	Specifically when $f_0=10^{-4.0} {\rm \ Hz}$, $\gamma_{\rm GD}$ is smaller than the reduction factor, $\gamma_{\rm acc} + \gamma_{\rm exp}$, and the effective growth rate is negative.
	The local maxima of $\gamma_{\rm GD}$ is determined by the balance between plasma beta and wave amplitude (see Eq. (\ref{eq:gamma_sagdeev})); the plasma beta is low and the wave amplitude is small in the lower corona and vice versa in the distant solar wind.

\subsection{Onset and suppression of PDI}

	\begin{figure}[!t]
		\begin{center}	 
 		 \includegraphics[width=70mm]{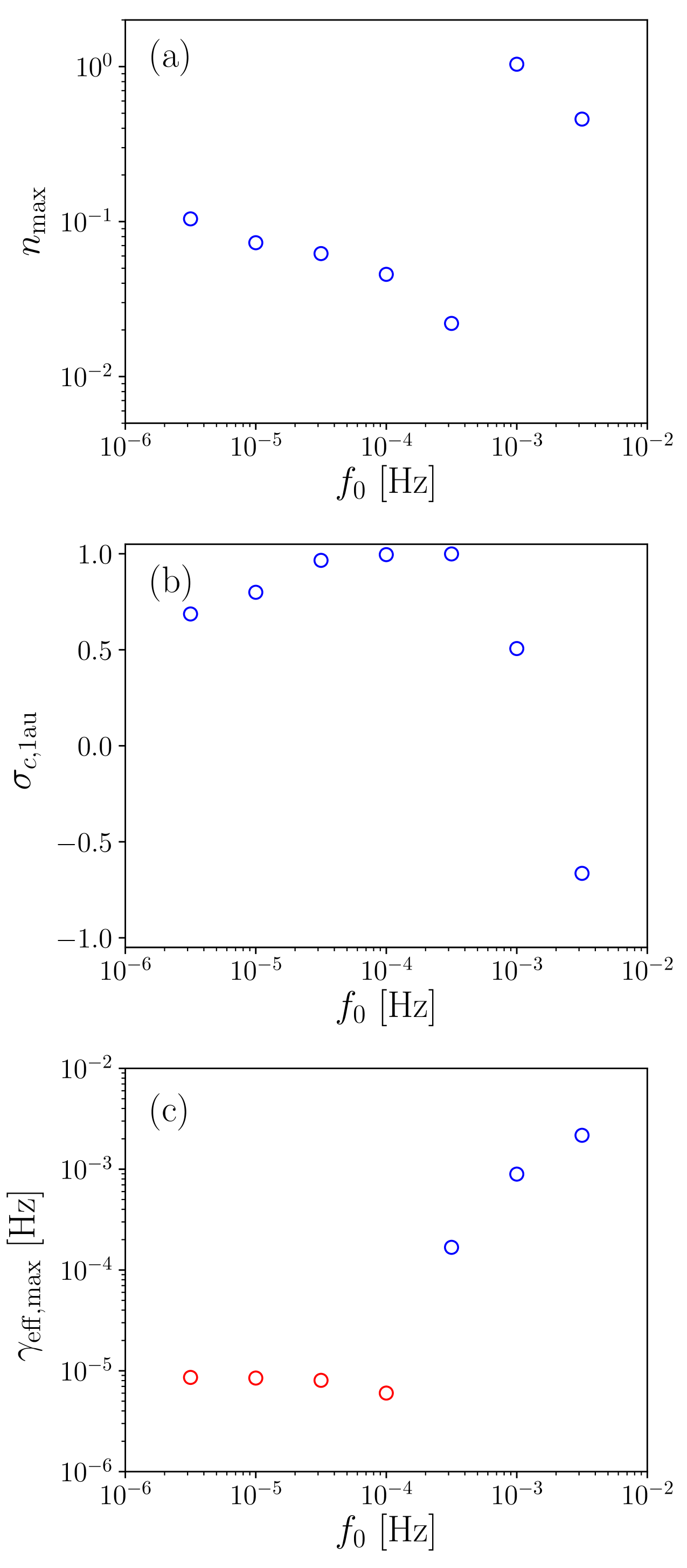} 
  		\end{center}
  		\vspace{-1em}
  		\caption{
			     Solar wind parameters versus wave-injection frequency $f_0$.
			     Each panel indicates
			     a: maximum fractional density fluctuation $n$,
			     b: normalized cross-helicity $\sigma_c$ at $1 {\rm \ au}$,
			     c: maximum effective growth rate $\gamma_{\rm eff}$ (blue: positive, red: negative).
	  				}
  		\label{fig:fdepend_drochgam}
	\end{figure}

	To discuss the threshold of the onset of PDI, we calculated the maximum fractional density fluctuation $n_{\rm max}$ and the normalized cross-helicity (Alfv\'enicity) $\sigma_c$ at $1 {\rm \ au}$.
	Here we defined $n_{\rm max}$ and $\sigma_c$ as
	\begin{align}
		n_{\rm max} &= \max \left( \frac{1}{\rho_{\rm ave}} \sqrt{\average{\left( \rho - \rho_{\rm ave} \right)^2}} \right), \ \ \ \ \rho_{\rm ave} = \average{\rho}, \\
		\sigma_c &= \frac{\average{z^{+}}^2 - \average{z^{-}}^2}{\average{z^{+}}^2 + \average{z^{-}}^2}, 
	\end{align}
	where $\average{X}$ denotes the time-averaged value of $X$ and $\max \left( X \right)$ denotes the maximum value of $X$ in space.
	We note that the sign of ${\sigma_c}$ is opposite to the sign of $H_c = \vec{v} \cdot \vec{B}$.
	These values can be useful indicators of PDI because PDI generates large-amplitude density fluctuation, which increases $n$, and back-scattered Alfv\'en waves, which reduce $\sigma_c$. 
	The latter effect works to reduce $\sigma_c$.
	According to \reft{Cranm12a}, without PDI, $\delta \rho_{\rm rms} / \rho_0 \lesssim 0.1$, and thus $\delta \rho_{\rm rms} / \rho_0 > 0.1$ indicates the PDI.
	
	In Figure \ref{fig:fdepend_drochgam}, we show a: $n_{\rm max}$, b: $\sigma_c$ at $1 {\rm \ au}$, and c: the maximum effective growth rate $\gamma_{\rm eff, max}$ (blue: positive, red: negative) as functions of $f_0$.
	When $f_0 \lesssim 10^{-3.5} {\rm \ Hz}$, both $n_{\rm max}$ and $\sigma_c$ show monotonic trends with $f_0$: 
	$n_{\rm max}$ decreases and $\sigma_c$ increases as $f_0$ increases.
	This is explained as follows.
	As $f_0$ becomes smaller, Alfv\'en waves are reflected more efficiently \refp{Ferra58,An00090,Velli93,Cranm05,Hollw07}.
	If Alfv\'en waves are reflected in the solar wind beyond the Alfv\'en point, reflected Alfv\'en waves are advected towards $1 {\rm \ au}$ and contribute to reducing $\sigma_c$.
	Note that the inward waves vanish near the Alfv\'en point \refp{Verdi09,Tener17a}.
	When the amount of reflected Alfv\'en waves increases, the interaction between outward and inward waves is activated.
	This wave-wave collision excites not only turbulence \refp{Irosh64,Kraic65,Dobro80,Goldr95}, 
	but also the slow-mode generation \refp{Wentz74,Uchid74} by the modulation of magnetic field pressure \refp{Hollw71,Kudoh99,Cranm15}.
	Magnetic field modulation also leads to direct steepening to fast shock \refp{Cohen74a,Kenne90,Suzuk04}.
	Owing to these processes, larger-density fluctuation is likely to be generated in the presence of larger-amplitude reflected Alfv\'en waves.
	
	\begin{figure}[!t]
		\begin{center}	 
 		 \includegraphics[width=70mm]{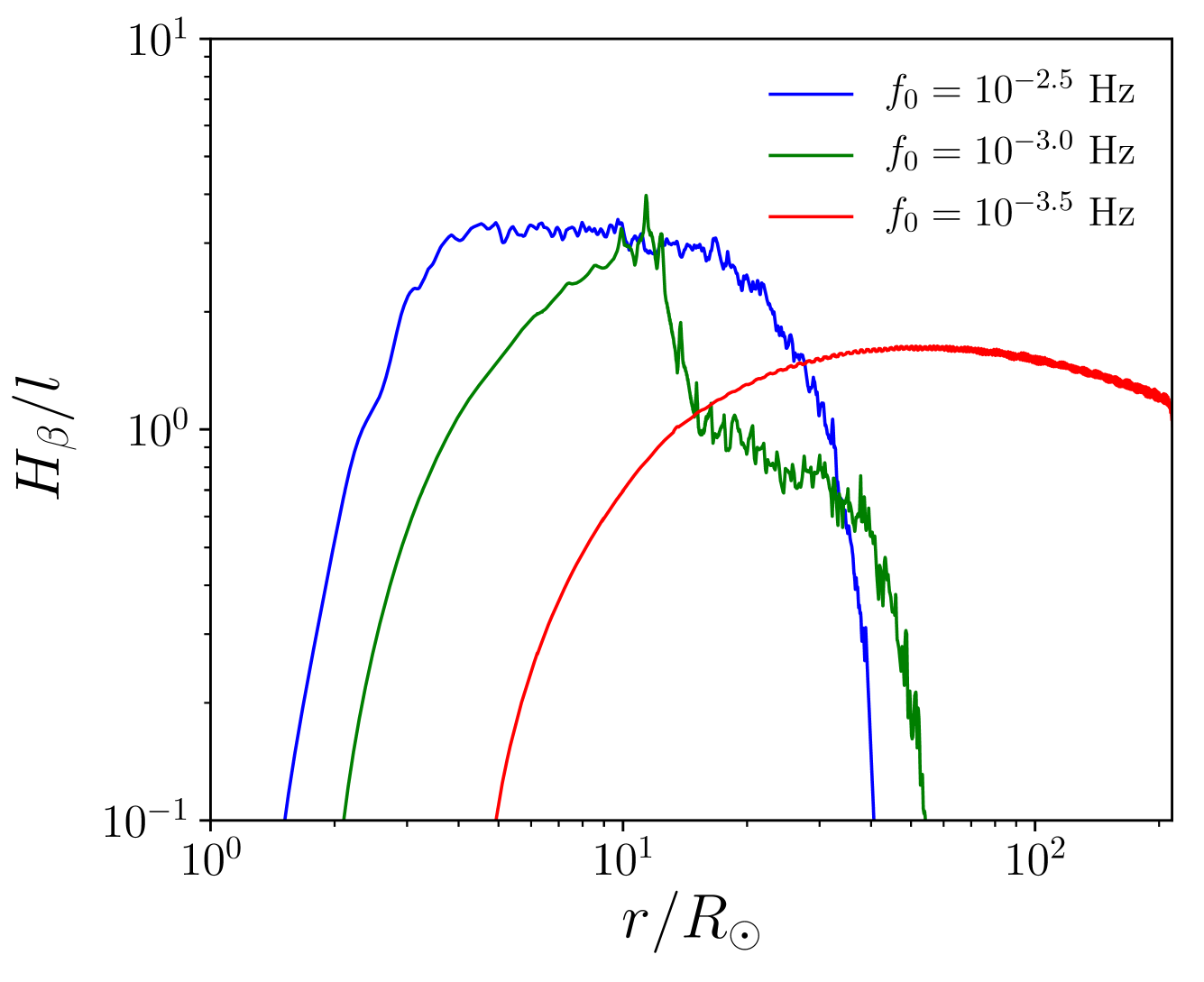} 
  		\end{center}
  		\vspace{-1em}
  		\caption{
			     Scale ratio $H_\beta / l$ 
			     where $l$ and $H_{\beta}$ denote the typical propagation length scale during the PDI growth and the scale length of plasma beta, respectively.
			     Blue, green, and red lines indicate $f_0 = 10^{-2.5} {\rm \ Hz}$, $10^{-3.0} {\rm \ Hz}$, and $10^{-3.5} {\rm \ Hz}$, respectively.
  				}
  		\label{fig:scaleratio}
	\end{figure}
	
	The monotonic trend in $10^{-5.5} {\rm \ Hz} \lesssim f_0 \lesssim 10^{-3.5} {\rm \ Hz}$ breaks down near $f_0 \approx 10^{-3} {\rm \ Hz}$.
	When $f_0$ gets larger than $10^{-3} {\rm \ Hz}$, $n_{\rm max}$ becomes larger than $0.1$ and $\sigma_c$ becomes smaller than $0.5$.
	Considering the fact that PDI generates large amounts of density fluctuation and reflected Alfv\'en waves, 
	Figure \ref{fig:fdepend_drochgam} indicates that the frequency threshold of the onset of PDI is $10^{-3.5} {\rm \ Hz} < f_0 < 10^{-3} {\rm \ Hz}$.
	This means that, 
	even though $\gamma_{\rm eff,max}$ is positive when $f_0 = 10^{-3.5} {\rm \ Hz}$,
	PDI cannot develop with this injection frequency.
	\reft{Tener13} argued that PDI is suppressed not only by the acceleration and expansion of the solar wind but also by the inhomogeneity of the solar wind, 
	because the resonance condition changes as the plasma parameters such as the plasma beta, Alfv\'en speed and wave amplitude, vary.
	In Figure \ref{fig:scaleratio}, we show the ratio between the propagation length during growth time $l$ and the scale length of the plasma beta ($H_\beta$):
	\begin{align}
		l = \frac{ v_A + v_r}{\gamma_{\rm eff}},  \ \ \ \ H_\beta = \left| \frac{\beta}{ \partial \beta / \partial r} \right|.
	\end{align}
	The ratio, $H_{\beta}/l$, can be used as a measure of how the inhomogeneity of the background field affects the onset of PDI;
	if $H_{\beta}/l$ is small $\lesssim 1$, the background inhomogeneity could suppress the PDI.
	Figure \ref{fig:scaleratio} shows $H_{\beta} / l$ versus height. 
	This indicates that, when $f_0 = 10^{-3.5} {\rm \ Hz}$, PDI cannot evolve because the scale ratio $H_{\beta} / l$ is at most around unity and the inhomogeneity affects the growth of PDI.
	
	Another possible reason that PDI is not observed when $f_0 = 10^{-3.5} {\rm \ Hz}$ is that the typical growth time is too small to develop before $1 {\rm \ au}$.
	$\gamma_{\rm eff}$ averaged over the entire simulation box is approximately $\overline{\gamma}_{\rm eff} = 5 \times 10^{-5} {\rm \ Hz}$, corresponding to the timescale of $\overline{\tau} = 2 \times 10 ^{4} {\rm \ s}$.
	Therefore, because it takes a few $\overline{\tau}$ to reach the saturation phase of PDI, 
	the evolution timescale ($\sim 10^{5} {\rm \ s}$) is comparable to the propagation timescale up to $1 {\rm \ au}$ ($\sim 2 \times 10^{5}  {\rm \ s}$).
	This indicates that $1 {\rm \ au}$ might be too short for the PDI of $f_0 = 10^{-3.5} {\rm \ Hz}$ to develop.
			
\section{Broadband-wave injection}  \label{sec:broad}

	Next, we applied the broadband-wave injection to discuss the consistency with observation.
	The boundary condition is given as
	\begin{align}
		z^{+}_{x,y} = \int^{f_{\rm max}}_{f_{\rm min}} P(f) \sin \left( 2 \pi f t + \phi_{x,y}(f) \right) df,
	\end{align}
	where $P(f)$ is determined so that the root-mean-square value is $2 v_{\rm rms,0}$ and the power show $1/f$ spectrum in $f_{\rm min} \le f \le f_{\rm max}$ \refp{Bruno13}.
	$\phi_{x,y}(f)$ are random functions of $f$.
	We fixed $f_{\rm max} = 10^{-2} {\rm \ Hz}$, corresponding to $100 {\rm \ s}$ in terms of period.
	Note that some observations show an even higher frequency component \refp{He00009,Okamo11,Shoda18b}.
	$f_{\rm min}$ is the free parameter.
	In this study, we calculated three cases: $f_{\rm min} = 10^{-3} {\rm \ Hz}$, $10^{-4} {\rm \ Hz}$, $10^{-5} {\rm \ Hz}$, 
	each of which corresponded to the largest timescale of 
	the photospheric transverse motion \refp{Matsu10b}, the coronal transverse motion \refp{Morto15}, and the solar-wind fluctuations \refp{Tu00095}, respectively.
	
	\begin{figure*}[!t]
		\begin{center}	 
 		 \includegraphics[width=170mm]{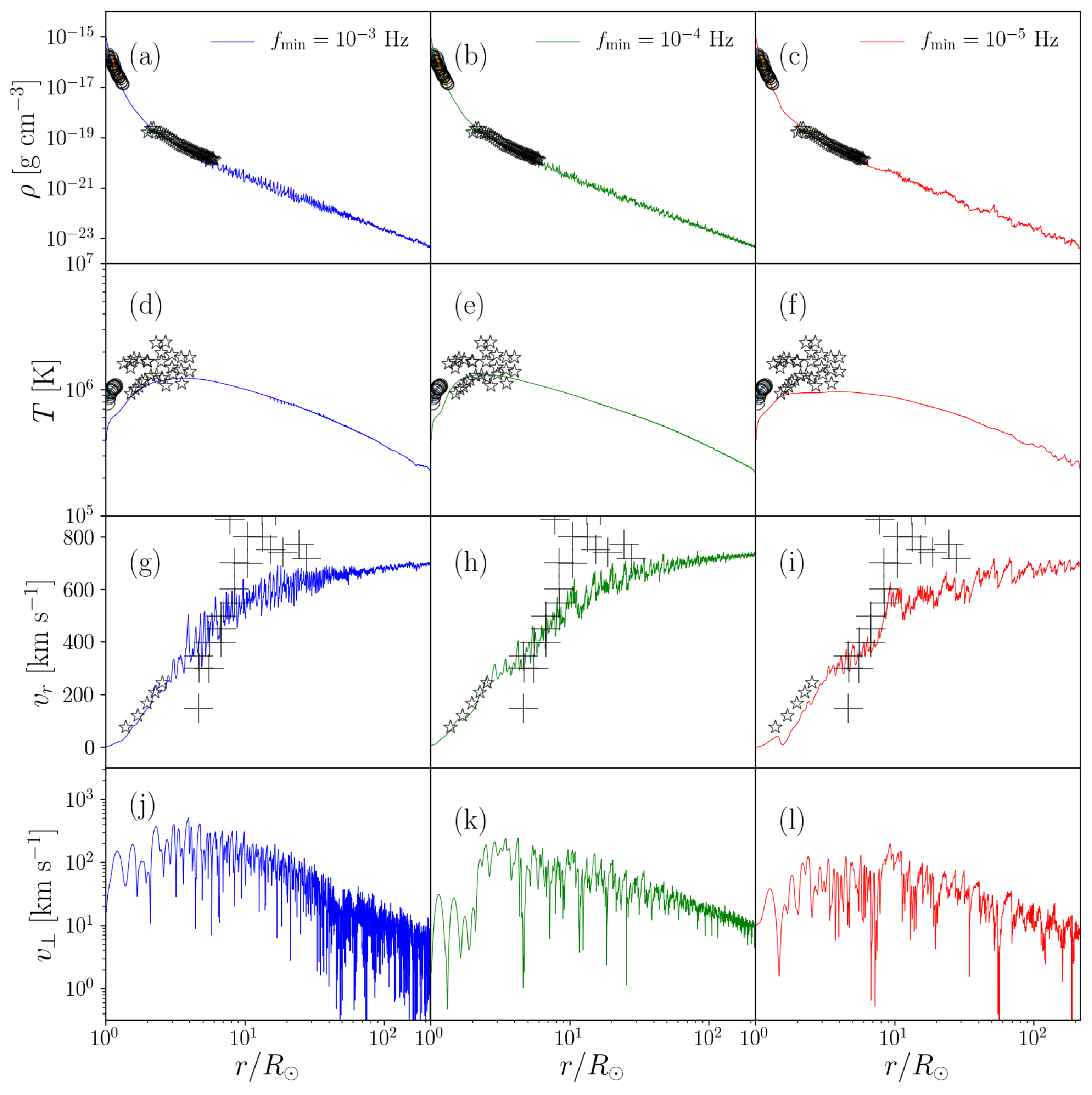} 
  		\end{center}
  		\vspace{-1em}
  		\caption{
				Quasi-steady states with different $f_{\rm min}$ values.
				Blue, green and red lines indicate $f_{\rm min} = 10^{-3} {\rm \ Hz}$, $10^{-4} {\rm \ Hz}$, $10^{-5} {\rm \ Hz}$, respectively.
				a-c: mass density. Shown by circles and stars are observations by \reft{Wilhe98} and by \reft{Lamy097}, respectively.
				d-f: temperature. Shown by circles and stars are observations by \reft{Landi08} and compilation of observed data by \reft{Cranm04,Cranm09b}, respectively.
				g-i: radial velocity. Shown by stars are ion outflow speed by \reft{Zangr02}, while the crosses represent the IPS observations \refp{Kojim04}.
				j-l: transverse velocity (Alfv\'en-wave amplitude).
  				}
  		\label{fig:compare_rs_broad}
	\end{figure*}

\subsection{Quasi-steady state}

	As in Section \ref{sec:mono}, we begin by discussing the quasi-steady states.
	Figure \ref{fig:compare_rs_broad} shows the same variables as those shown in Figure \ref{fig:compare_rs_mono} 
	except for Panel d, where the transverse velocity $v_\perp$ is shown instead of the Els\"asser variables.
	Color represents $f_{\rm min} = 10^{-3} {\rm \ Hz}$ (blue), $10^{-4} {\rm \ Hz}$ (green), $10^{-5} {\rm \ Hz}$ (red), respectively.
	Because the main motivation of broadband-wave injection is to compare to observations, we also show several observational values.
	In Panel a, we show the density observation by \reft{Wilhe98} (circles) and \reft{Lamy097} (stars).
	In converting the observed electron density $n_e$ to mass density $\rho$, we simply assumed $\rho = m_p n_e$.
	In Panel b, circles and stars correspond to the results from \reft{Landi08} and \reft{Cranm04,Cranm09b}, respectively.
	In Panel c, observed ion-outflow velocity is plotted by stars \refp{Zangr02}, while the results of IPS observation are indicated by crosses \refp{Kojim04}.

	\begin{figure}[!t]
		\begin{center}	 
 		 \includegraphics[width=70mm]{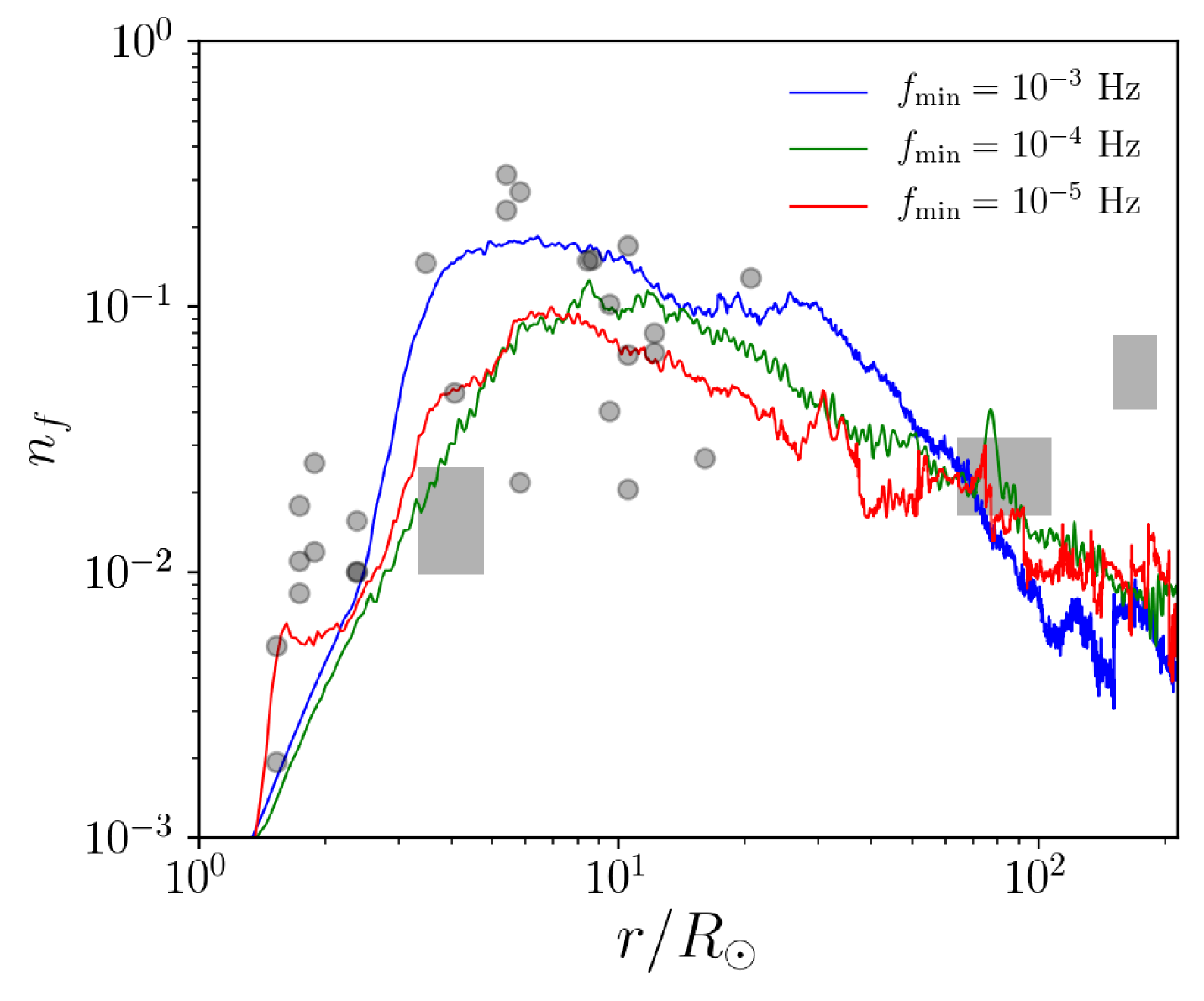} 
  		\end{center}
  		\vspace{-1em}
  		\caption{
				Fractional density fluctuation $n_f$ versus heliocentric distance.
			 	Blue, green and red lines indicate $f_{\rm min} = 10^{-3} {\rm \ Hz}$, $10^{-4} {\rm \ Hz}$, $10^{-5} {\rm \ Hz}$, respectively.
				Shown by gray rectangles and circles are the observational values (see text).
  			     }
  		\label{fig:freq_dro}
	\end{figure}
	
\subsection{Density fluctuation}

	The density fluctuation in the solar wind is observed by radio-wave observation.
	As explained in the Introduction, density fluctuation possibly plays a role in reflecting Alfv\'en waves in the corona and solar wind \refp{Balle16}.
	
	When we applied the broadband-wave injection, it was difficult to obtain the amplitude of the density fluctuation that was solely attributed to PDI,
	because the density fluctuates not only by PDI but also by the time variation of the injected energy flux.
	Since the density fluctuation that comes from the injection has a timescale of typically $f_{\rm min}$, 
	in this study, we defined the density fluctuation as high-frequency ($>f_{\rm min}$) components.
	Given that we have density $\rho(r,t)$, the fractional fluctuation $n_f (r)$ is given as
	\begin{align}
		n_f (r) = \frac{1}{\rho_{\rm ave}} \sqrt{ \frac{1}{2 \pi \tau_0} \int_{\left| \omega \right| > 2 \pi f_{\rm th}} \left| \tilde{\rho} (r,\omega) \right|^2 d\omega},
	\end{align}
	where
	\begin{align}
		\tilde{\rho} (r,\omega) = \int^{\tau_0}_0 dt \rho (r, t) e^{ i \omega t}.
	\end{align}
	Note that Parseval's identity holds as follows:
	\begin{align}
		\int dt \left| \rho(r,t) \right|^2 = \frac{1}{2 \pi} \int d \omega \left| \tilde{\rho} \left(r, \omega \right)  \right|^2.
	\end{align}
	$f_{\rm th}$ is the frequency threshold.
	Here, we set $f_{\rm th} = 10^{-3} {\rm \ Hz}$.
	Although this value was a rather arbitrary choice, 
	we confirmed that the radial trend of $n_f$ does not depend on $f_{\rm th}$.
	
	Figure \ref{fig:freq_dro} shows the radial profiles of $n_f (r)$.
	Rectangles are observational values taken from \reft{Cranm12a}.
	The rectangle near $r = 5 R_\odot$ indicates the radio sounding data \refp{Coles89,Spang02,Harmo05} while the rectangles in $r \gtrsim 100 R_\odot$ indicate the in-situ data by \reft{Marsc90}.
	The circles indicate the observation by \reft{Miyam14}.
	
	Our three cases nicely explain the overall radial profile of the observed density fluctuation peaked at $r\sim 5-10 R_{\odot}$ \refp{Miyam14}.
	The peak of $n_f$ in our calculation is created by the high-frequency ($f>10^{-3.5}$Hz) Alfv\'{e}n waves that are subject to PDI (Fig.3b). 
	The largest effective growth rate $\gamma_{\rm eff}$ is peaked in $r \sim 2-10 R_\odot$ when the parent-wave frequency is $10^{-3} - 10^{-2} {\rm \ Hz}$.
	Therefore, the large density fluctuations are excited as an outcome of the PDI in these locations. 
	To summarize, the observed density fluctuation is explained by the evolution of the PDI of high-frequency ($10^{-2} - 10^{-3} {\rm \ Hz}$) Alfv\'en waves.
	
\subsection{Cross-helicity in the solar wind}

\begin{figure}[!t]
		\begin{center}	 
 		 \includegraphics[width=70mm]{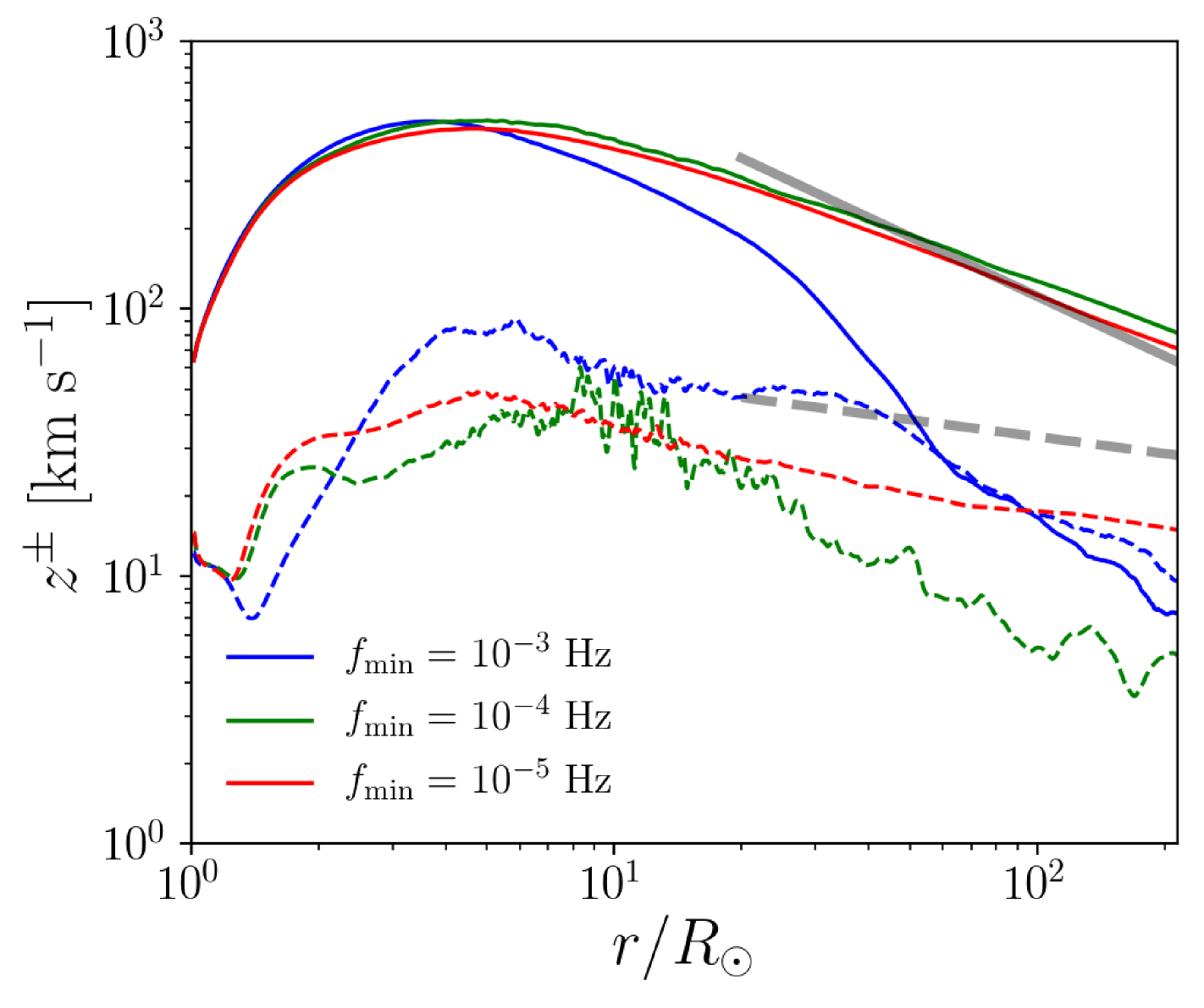} 
  		\end{center}
  		\vspace{-1em}
  		\caption{
				Radial profiles of time-averaged Els\"asser variables.
				Solid and dashed lines indicate $z^{+}$ (anti-sunward) and $z^{-}$ (sunward) components.
				Blue, green and red lines indicate $f_{\rm min} = 10^{-3} {\rm \ Hz}$, $10^{-4} {\rm \ Hz}$, $10^{-5} {\rm \ Hz}$, respectively.
				Also shown by gray lines are observational trends by \reft{Bavas00}.
  				}
  		\label{fig:els_compare}
	\end{figure}

	Because the radial evolution of Els\"asser variables $z^{\pm}$ in the heliosphere has been observed \refp{Bavas82,Bavas00}, we can test our theoretical model by comparison with these observations.
	In Figure \ref{fig:els_compare}, we show the radial profile of time-averaged Els\"asser variables ($z^{+}$: solid line, $z^{-}$: dashed line) with different $f_{\rm min}$ values
	($f_{\rm min}=10^{-3} {\rm \ Hz}$: blue, $10^{-4} {\rm \ Hz}$: green, $10^{-5} {\rm \ Hz}$: red).
	Also shown by gray transparent lines are the observational trends by \reft{Bavas00}.
	
	While $z^{+}$ is consistent with observation when $f_{\rm min}=10^{-4}, \ 10^{-5} {\rm \ Hz}$, we have a much smaller $z^{+}$ compared with observation when $f_{\rm min}=10^{-3} {\rm \ Hz}$.
	Because PDI evolves when $f_0 \gtrsim 10^{-3} {\rm Hz}$, this discrepancy indicates that, via PDI, excessive energy transfer from $z^{+}$ to $z^{-}$ occurs.
	When $f_{\rm min}$ becomes smaller, the intensity of high-frequency waves that are subject to PDI is reduced because the total wave power is fixed.
	This is why the signature of PDI is weak for smaller $f_0$.
	Our result indicates that the cross-helicity evolution in the solar wind is dominated by the linear reflection \citep{Zhou090,Velli91,Verdi07}.
	Because the simulated $z^{-}$ approaches the observational value as $f_{\rm min}$ decreases, as a result of PDI suppression, 
	the cross-helicity evolution in the solar wind is governed by linear reflection of low-frequency ($f_0 \sim 10^{-5} {\rm \ Hz}$) components.

\section{Summary \& Discussion} \label{sec:summary}
	
	In this study, using numerical simulations, we investigated the threshold of the onset of PDI by changing the Alfv\'en-wave injection.
	As discovered by \reft{Tener13} and \reft{DelZa15}, wind acceleration and expansion work to reduce the growth rate of PDI.
	We have solved the wave propagation self-consistently from the coronal base to $1 {\rm \ au}$, and this was then compared with the accelerating expanding box simulation.
	
	Firstly, we investigated the fundamental processes of PDI by applying monotonic-wave injection with frequency $f_0$.
	Our results show that PDI can develop when $f_0 \gtrsim 10^{-3} {\rm \ Hz}$, while we observe no signature of PDI when $f_0 \lesssim 10^{-3.5} {\rm \ Hz}$.
	Owing to the wind acceleration and expansion, the growth rate of PDI becomes negative when $f_0 \lesssim 10^{-4.0} {\rm \ Hz}$.
	When $f_0 \lesssim 10^{-3.5} {\rm \ Hz}$, even though the growth rate of PDI is positive, PDI cannot develop.
	The suppression by solar wind inhomogeneity or the long timescale of growth might be the reason for this.	
	
	The frequency-filtering mechanism can operate in the corona and solar wind due to the bimodal behavior of wave dissipation with respect to frequency.
	The low-frequency ($f_0\lesssim 10^{-4} {\rm \ Hz}$) waves undergo linear reflection and generate Alfv\'{e}nic turbulence from the interaction with counter-propagating waves. 
	The high-frequency ($f_0\gtrsim 10^{-3} {\rm \ Hz}$) waves dissipate through the PDI. As a result of the efficient heating, dense, hot and relatively slow winds are driven in the cases with $f_0\lesssim 10^{-4} {\rm \ Hz}$ or $f_0\gtrsim 10^{-3} {\rm \ Hz}$. 
	In contrast, the intermediate-frequency ($f_0\approx 10^{-3.5} {\rm \ Hz}$) waves are not severely subjected to these damping mechanisms.
	As a result, fast and less dense wind emanates from the relatively cool corona in this case. 
	This indicates that the corona and solar wind have a frequency-filtering effect of the Alfv\'en wave, and as a result, the medium-frequency wave is likely to permeate.
	This is a possible reason for the hour-scale waves observed in the solar wind \citep{Belch71a}.
	
	Secondly, we applied broadband-wave injection to compare the numerical results with observation.
	The observed radial trend of the density fluctuation can be well explained by the evolution of the high-frequency ($f_0\gtrsim 10^{-3} {\rm \ Hz}$) Alfv\'en waves.
	However, the observed trend of the cross-helicity can be explained by the linear reflection of the low-frequency ($f_0 < 10^{-4} {\rm \ Hz}$) Alfv\'en waves.
	These results show that the Alfv\'en waves in a wide range of frequency play an essential role in the global solar wind. 
	
	There are several limitations in our model.
	The most severe limitation is the treatment of turbulence.
	We have applied a simple one-point-closure model of Alfv\'en wave turbulence (Eq. (\ref{eq:eta1}) and (\ref{eq:eta2})) with the correlation length that increases with an expanding flux tube (Eq. (\ref{eq:lambda})).
	However, \reft{Cranm12a} showed that Eq. (\ref{eq:lambda}) possibly underestimates the correlation length.
	To overcome this, we needed to solve the transport equation of $\lambda$ \refp{Breec08,Usman11}.
	In addition, the correlation length should be different between $z^{+}$ and $z^{-}$ \refp{Zank017,Shiot17}.
	More sophisticated treatment of the turbulence, including the shell model \refp{Buchl07,Verdi12a}, remains as a future work.
	
	Another limitation is one-dimensional modeling.
	While the Alfv\'en wave turbulence is taken into account phenomenologically, we completely ignore the effect of phase mixing \refp{Heyva83,Kanek15,Antol15,Okamo15} by 1D modeling.
	Besides, it has been shown by \reft{DelZa01} that the onset (and possibly growth) of parametric decay instability is slower in 3D than in 1D.
	Our quantitative discussion might be slightly modified by the 1D assumption.
	Also, we cannot take into account the wave refraction in the lower region \refp{Rosen02,Bogda03}.
	In future, we need to conduct a 3D MHD simulation for the reasons above.
	
	In this study, we have focused on the frequency dependence.
	Since the growth rate of parametric decay instability depends also on plasma beta and wave amplitude \refp{Golds78,Derby78}.
	\reft{Suzuk06a} investigated the dependence on the injected wave amplitude. 
	Readers probably expect that the PDI is suppressed for smaller wave injection according to Eq. (\ref{eq:gamma_sagdeev}). 
	However, the response of the solar wind totally changes the situation. 
	A case with smaller injection gives lower coronal temperature because of the suppressed heating. 
	As a result, the plasma beta in the corona is lower, and larger density fluctuations are excited by more activated PDI as shown in Figure 9 of \reft{Suzuk06a}.
	Similarly, it is expected that the density variation is large when the magnetic field is stronger and the plasma beta is lower.
	
	There are also ambiguities in the thermal flux in the free-streaming regime. 
	We chose $\alpha=2$ in evaluating the magnitude of free-streaming thermal flux.
	Although $\alpha=4$ has been sometimes used \refp{Leer082,Withb88,Landi03,Cranm07,Balle16},
	$\alpha=4$ might overestimate the actual flux because thermal conduction can be suppressed by the local instability and turbulence \refp{Gary099,Rober17,Komar17,Tong018}.
	Indeed \reft{Cranm09a} showed that $\alpha=1.05$ yields good agreement with observation, and several recent studies used $\alpha=1.05$ \refp{Usman11,Holst14}.
	The precise value of $\alpha$ should depend on the solar wind condition.
	Since the change in $\alpha$ does not strongly affect the physical quantities of the solar wind \refp{Cranm07}, we expect that our findings are independent on $\alpha$.

	M.S. is supported by the Leading Graduate Course for Frontiers of Mathematical Sciences and Physics (FMSP) and Grant-in-Aid for Japan Society for the Promotion of Science (JSPS) Fellows.
	T.Y. is supported by JSPS KAKENHI Grant Number 15H03640.
	T.K.S. is supported in part by Grants-in-Aid for Scientific Research from the MEXT of Japan, 17H01105.
	Numerical calculations were in part carried out on the PC cluster at the Center for Computational Astrophysics, National Astronomical Observatory of Japan.

\end{document}